\documentclass[english]{article}
\usepackage[T1]{fontenc}
\usepackage[latin9]{inputenc}
\usepackage{amsbsy}
\usepackage{amssymb}
\usepackage{graphicx}
\usepackage[numbers]{natbib}

\makeatletter

\providecommand{\tabularnewline}{\\}

\usepackage{amsmath,epsfig}
\usepackage{comment}
\usepackage{amsthm}

\usepackage{algorithm,algpseudocode}

\title{Characterising Dependency in Computer Networks using Spectral Coherence}

\author{Alex J. Gibberd$^*$, Jordan Noble, Edward A.K. Cohen\thanks{Funded by EPSRC grant EP/P011535/1}}

\makeatother

\usepackage{babel}
\begin{document}

\maketitle
\begin{abstract}
The transmission or reception of packets passing between computers
can be represented in terms of time-stamped events and the resulting
activity understood in terms of point-processes. Interestingly, in
the disparate domain of neuroscience, models for describing dependent
point-processes are well developed. In particular, spectral methods
which decompose second-order dependency across different frequencies
allow for a rich characterisation of point-processes. In this paper,
we investigate using the spectral coherence statistic to characterise
computer network activity, and determine if, and how, device messaging
may be dependent. We demonstrate on real data, that for many devices
there appears to be very little dependency between device messaging
channels. However, when significant coherence is detected it appears
highly structured, a result which suggests coherence may prove useful
for discriminating between types of activity at the network level.
\end{abstract}

\section{Introduction}

Understanding how devices on computer networks communicate is a challenging
task. While it is possible to gather vast quantities of data from
such networks, for instance via packet monitoring, it is difficult
to store, let alone process. As a result, protocols such
as NetFlow which sample and summarise packet level data are now very
popular \citep{Duffield2004}. Even still, regular monitoring protocols
can produce hundreds of gigabytes of summary statistics on a network
per day which need to be converted into actionable insights for network
administrators.

Network defenders should be at a theoretical advantage over attackers,
in that they can attempt to model and understand the day-to-day activity
of their network. From such models they can then define what anomalous,
and/or malicious events may look like. Additionally, in order to enhance
detection performance, one may desire to use prior knowledge of what
benign network activity should look like in order to define anomalies.
For example, and relevant to the approach developed here, one may
expect that communication between pairs of devices are not correlated
such that their activity should be broadly independent when monitored
at a network level. However, if traffic is dependent across
pairs then this may indicate potentially malicious behaviour 
such as lateral movement or tunnelling \cite{Neil2013}.

There are a great variety of measures and methods that can be used
to analyse dependency between streams, for instance through measures
such as covariance \citep{Jin2007}, correlation \cite{Neil2013}, partial correlation
\citep{Gibberd2017}, or higher-order measures such as cross-cumulants
\citep{Brillinger}. A traditional approach to
network traffic modelling is to assume it is generated according to
a Poisson point process \citep{Duffield2004,Murdoch2007}. While such
models may be generalised to a multivariate setting \citep{Baurele2005},
they do not allow for us to encode auto-correlation structure within
a point-process. As a result, we may be able to describe processes
which are dependent on each other across data-streams, but they do
not allow for dependency within a data-stream itself. When considering
computer network traffic, it is not hard to imagine that events from
a device will be somehow dependent on previous events from that same
device. Spectral approaches, based on either Fourier \citep{Brillinger1981}
or time-scale wavelet analysis \citep{Riedi1999,Scherrer2007}
of processes provide a valuable tool in this situation as they allow
for both a rich description of auto-correlation and cross-channel
dependency \citep{Cohen2014}. 

In this paper, we propose to utilise a measure known as spectral coherence
\citep{Carter1987} to characterise dependency between network communication
channels. We are not aware of any previous application of such a measure
to network traffic analysis, although the method has received great
attention in neuroscience for modelling neuron dependency \citep{Jarvis2001}. 

\section{Dataset and Preprocessing}

Consider network connected devices $A,B,C$ and their associated users,
for example, these may be personal computers, DNS servers, authentication
servers, or even printers. Typically, we would expect these devices
to go about their work as fairly independent actors, i.e. they may
browse websites, download material etc., but not in any particularly
coordinated manner. Device communication is typically performed through
packet transmission. However, given network monitoring limitations,
the events that we analyse need not necessarily be packets themselves. More likely, they are aggregates or summaries of communication, for
example NetFlow sessions.

\begin{figure}[t]
\begin{center}
\includegraphics[width=\columnwidth]{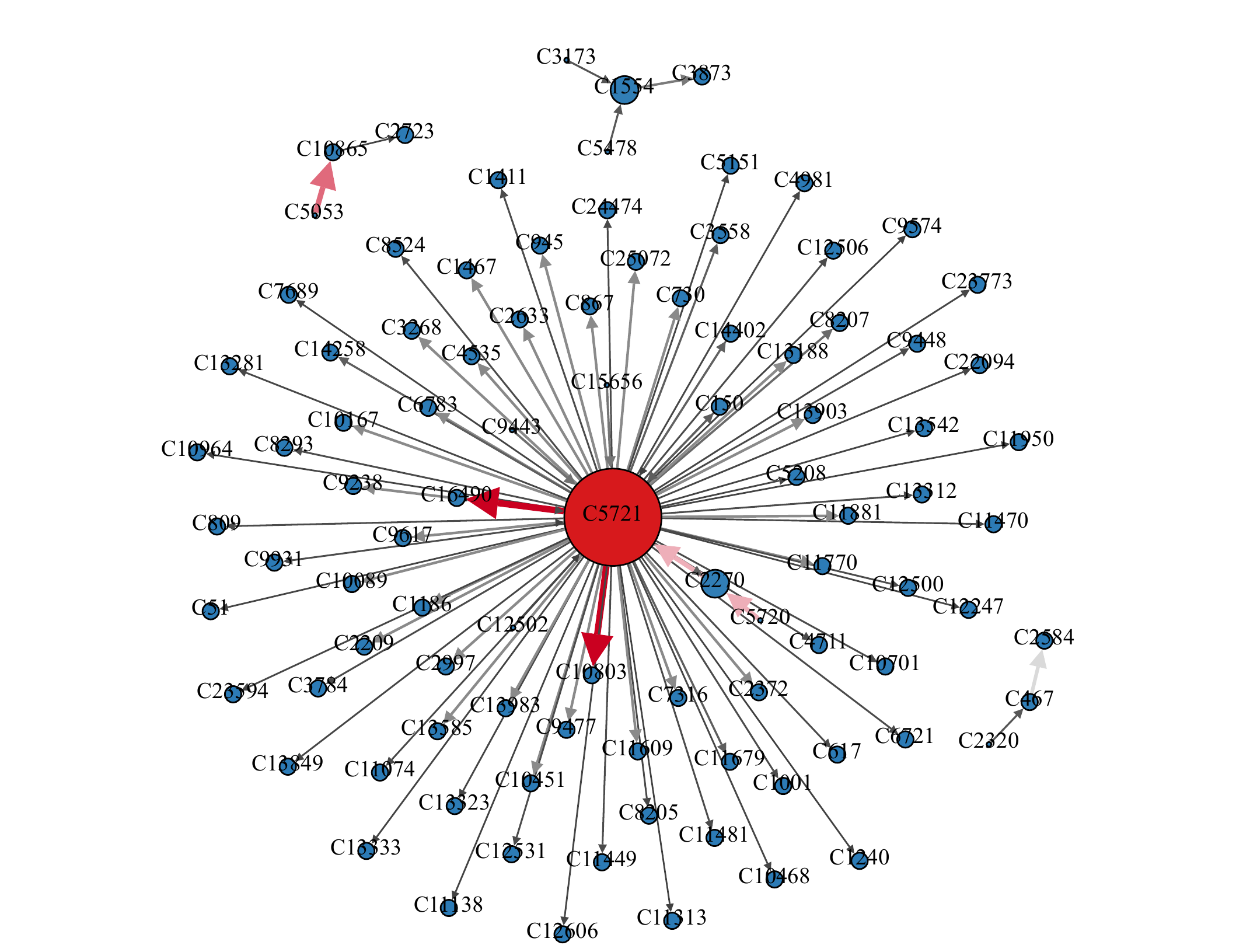}
\par\end{center}%
\begin{center}

\begin{tabular}{|c|c||c|c|}
\hline 
{\small{}$N_{\mathrm{triple}}$} & {\small{}500} & {\small{}Average length $T$} & {\small{}17.2 (hours)}\tabularnewline
\hline 
{\small{}Number of nodes} & {\small{}95} & {\small{}Average rate $\lambda_{AB}$} & {\small{}0.101 (1/s)}\tabularnewline
\hline 
{\small{}Unique edges} & {\small{}96} & {\small{}Average rate $\lambda_{BC}$} & {\small{}0.032 (1/s)}\tabularnewline
\hline 
\end{tabular}

\end{center}

\caption{Graph of messaging channels under analysis. Size of text and node
are respectively proportional to the out-going and in-going degree.
The weight of the edges (and colour) is proportional to the rate measured
on that edge.\label{fig:Graph_of_edge_pairs}}
\end{figure}

In our case, we analyse a subset of NetFlow session data from the
Los Alamos National Laboratory (LANL) multi-source cyber-security
events data \citep{kent-2015-cyberdata1,akent-2015-enterprise-data}.
More specifically, we create a subset of events (NetFlow session start
times) relating to the top $N_{\mathrm{triple}}=500$ busiest edge-pairs
in the network over a single day's (Thursday) worth of data. We assess
dependency in a pair-wise manner such that data-streams correspond to 
directed edge-communication between devices $A\rightarrow B$ and $B\rightarrow C$
for $i=1,\ldots,N_{\mathrm{triple}}$ device triples $(A,B,C)_{i}$. 
For each triple, the activity for
each edge corresponds to the same time frame. The protocol monitored is the same
for all edge pairs.
To avoid confusion, we exclude all flows from the triple $(C,B,A)$ when the triple
$(A,B,C)$ is included. One should note that our selection criteria for
data-set construction does not explicitly specify devices which have
a particular function on the network. However, if we look at the graph
of communication edges in Figure \ref{fig:Graph_of_edge_pairs} it
appears that many of our triples have repetitive edges, there
are only $95$ unique devices in our data-set and $96$ unique edges.
Looking at the topology of the network it appears that most of the
devices are communicating through the device C5721, while we do not
have labelled data relating to the function of devices, it would appear
that this node acts as some form of server.

We note that in our recordings it is possible to observe two events
which have the same start time. This means that the events cannot reasonably be treated
as being observed in continuous time, and indeed the timestamps provided
with our data are only accurate to the second. As such, the raw events
are aggregated into bins of width $\Delta=1$s. The binned bivariate process
will be denoted $\{\boldsymbol{X}[k]  = [X_{AB}[k],X_{BC}[k]]^T; k\in\mathbb{Z} \}$, for which we observe a portion 
$\boldsymbol{X}[1], \boldsymbol{X}[2],...,\boldsymbol{X}[K]$.
As a pre-processing step, we subtract the empirical mean of
the data-streams, relabeling $\boldsymbol{X}[k]:=\boldsymbol{X}[k]-\bar{\boldsymbol{X}}$
so that they can be well approximated as zero-mean processes.

\section{Spectral Coherence as a Measure of Association}

In this section, we will define the spectral coherence as a property
relating to the cross-spectrum of a bivariate process. The discussion
here will be developed based on the understanding that we are with
observations relating to a discrete-time process. However, it is also
possible to perform such analysis at the individual event level, for
examples, see the work of \citep{Cohen2014,Jarvis2001,Brillinger1972}. 

To simplify the notation, let us use $X_{1}[k]\equiv X_{AB}[k]$
and $X_{2}[k]\equiv X_{BC}[k]$.
Furthermore, we will assume that $\{X_{1}[k],X_{2}[k]\}$
represent a jointly second-order stationary process, i.e. the covariance $\Sigma_{ij}[\tau]\equiv\mathrm{Cov}(X_{i}[k+\tau],X_{j}[k])$
only depends on $\tau$ for $i,j=1,2$. Provided
$\sum_{\tau}|\Sigma_{ij}[\tau]|<\infty$, then for all frequencies $|\omega|\le \pi/\Delta$, the function $S_{ij}(\omega)=\Delta\sum_{\tau=-\infty}^{\infty}\Sigma_{ij}[\tau]e^{-i\omega\tau\Delta}$, $i,j=1,2$, is termed the spectum of $\{X_i[k]\}$ when $i=j$, and the \emph{cross-spectrum} between $\{X_{1}[k]\}$ and $\{X_{2}[k]\}$ when $i\neq j$. These spectra can be conveniently represented with the spectral matrix $\boldsymbol{S}(\omega) = (S_{ij}(\omega))$. The argument
we present in this paper, is that the cross-spectrum provides a rich
framework within which we may characterise computer network messaging
processes. In particular, since we are interested in dependency between
data-streams, we will concern ourselves with the squared coherencey, or ordinary coherence,
defined as the real-valued quantity
\begin{equation}
R(\omega)=\frac{|S_{12}(\omega)|^{2}}{S_{11}(\omega)S_{12}(\omega)}\;.\label{eq:Coherence_def}
\end{equation}
The coherence provides a useful statistic for assessing dependency
between point-processes; not only does it permit a decomposition over
frequencies allowing one to highlight periodicities associated with
dependence, but is also invariant to scaling of the marginal auto-covariance
as the measure is normalised by the on-diagonal spectra.

\subsection{Estimating the Spectra of Point-Processes}

Since the true values of the spectra, cross-spectra and coherence are unknown to
us, we are required to estimate them from data. The approach that
we utilise here is based on the work of Thomson \citep{Thomson1982}. Specifically, we will construct our estimators
from the \emph{tapered discrete-time Fourier transform (tDFT)} defined
as
\begin{equation}
\hat{F}_{j;l}(\omega)\equiv\Delta^{1/2}\sum_{k=1}^{K}h_{l}[k]X_{j}[k]e^{-i\omega k\Delta}\quad j=1,2,\label{eq:raw_DFT}
\end{equation}
for frequencies $-\pi/\Delta<\omega<\pi/\Delta$ where $\{h_{l}[k];k=1,...,N\}$
for $l=1,\ldots,L$ are a set of taper sequences. The tapers in the
above construction are important, in that they enable us to selectively
transform data-points prior to taking the Fourier transform.

If we temporarily assume that $h_{l}[k]=1$ for all $l,k$, then taking
the conjugate outer-product leads to the periodogram $\hat{I}_{ij;l}(\omega)\equiv \hat{F}_{i;l}(\omega)\hat{F}_{j;l}^{*}(\omega)$.
Unfortunately, while the periodogram is an asymptotically unbiased
estimator of the spectrum $E[\hat{I}_{ij;l}(\omega)]\rightarrow S_{ij}(\omega)$
as $T\rightarrow\infty$, it is not consistent, in that $\mathrm{Var}[\hat{I}_{ij;l}(\omega)]\not\rightarrow0$.
Principally, this is due to us attempting to estimate the spectra
at an infinite number of frequencies $\omega\in\mathbb{R}$ with only
a finite portion of data \citep{Brillinger1972,Jarvis2001}.

There are several approaches which can be used to sculpt asymptotically
consistent estimators of the spectra \citep{Nuttall1982,Thomson1982,Walden2000}.
A general strategy \citep{Walden2000}, is to adapt the direct spectral
estimate (where $h_{l}[k]=1$) such that the tapers take different
shapes, for example they may be supported in disjoint regions \citep{Brillinger1981},
or constitute a set of overlapping windows \citep{Nuttall1982}. From
the Fourier transform of the tapered data, we may then construct vectors $\hat{\boldsymbol{F}}_{l}(\omega)=[\hat{F}_{1;l},\hat{F}_{2;l}]^T$ and compute
what is known as a \emph{multi-taper spectral estimator} by averaging:
\[
\hat{\boldsymbol{S}}(\omega)=\frac{1}{L}\sum_{l=1}^{L}\hat{\boldsymbol{F}}_{l}(\omega)\hat{\boldsymbol{F}}_{l}^{H}(\omega)\quad i,j=1,2,
\]
where $^H$ denotes the complex conjugate transpose. From the multi-taper spectral estimate we may then obtain an estimate
for the coherence $\hat R_{12}(\omega)$ via (\ref{eq:Coherence_def})
replacing the true spectra $S_{11}(\omega)$, $S_{22}(\omega)$ and $S_{12}(\omega)$ with
the estimates $\hat S_{11}(\omega)$, $\hat S_{22}(\omega)$ and $\hat S_{12}(\omega)$, respectively.

\subsection{Taper Specification}

If we consider choosing orthogonal taper sequences whereby $\sum_{k}h_{l}[k]h_{l'}[k]=0$
for $l\ne l'$ then the resultant Fourier transforms will be asymptotically
independent \citep{Brillinger1981}. Averaging over these independent
sequences can then reduce the variation in the estimate, the reduction
will be related to the number of tapers we average over. It has been
demonstrated, c.f. \citep{Brillinger1981}, that in the case of asymptotically
orthogonal tapers, the sampling distribution of the spectral matrix $\hat{\boldsymbol{S}}(\omega) = (\hat S_{ij}(\omega))$ is given
by a 2D complex Wishart distribution $\hat{\boldsymbol{S}}(\omega)\sim W_{2}^C(L,\boldsymbol{S}(\omega))$
with $L$ degrees of freedom and scale matrix $\boldsymbol{S}(\omega)$. 

In our application, we utilise a form of taper first demonstrated
for spectral estimation by Thomson \citep{Thomson1982}. Often referred
to as the Slepian tapers, these sequences have the beneficial properties
that they are mutually orthogonal while maximising
energy concentration in a small frequency interval $[-\omega_{W},\omega_{W}]$.
If two frequencies are separated by more than this bandwidth, then
the bias due to tapering is in some sense minimised. However, as the
number of tapers $L$ increases, the width of the side-lobe associated
with the Fourier transform of $h_{L}[k]$ necessarily increases. As
such, there is a classic bias-variance trade-off, increasing $L$
reduces the variance, but increases bias. The appropriate number of
tapers to use is highly dependent on application and something we
will shortly revisit in the context of the network traffic dataset.

\begin{figure}[t]
\begin{centering}
\includegraphics[width=1\columnwidth]{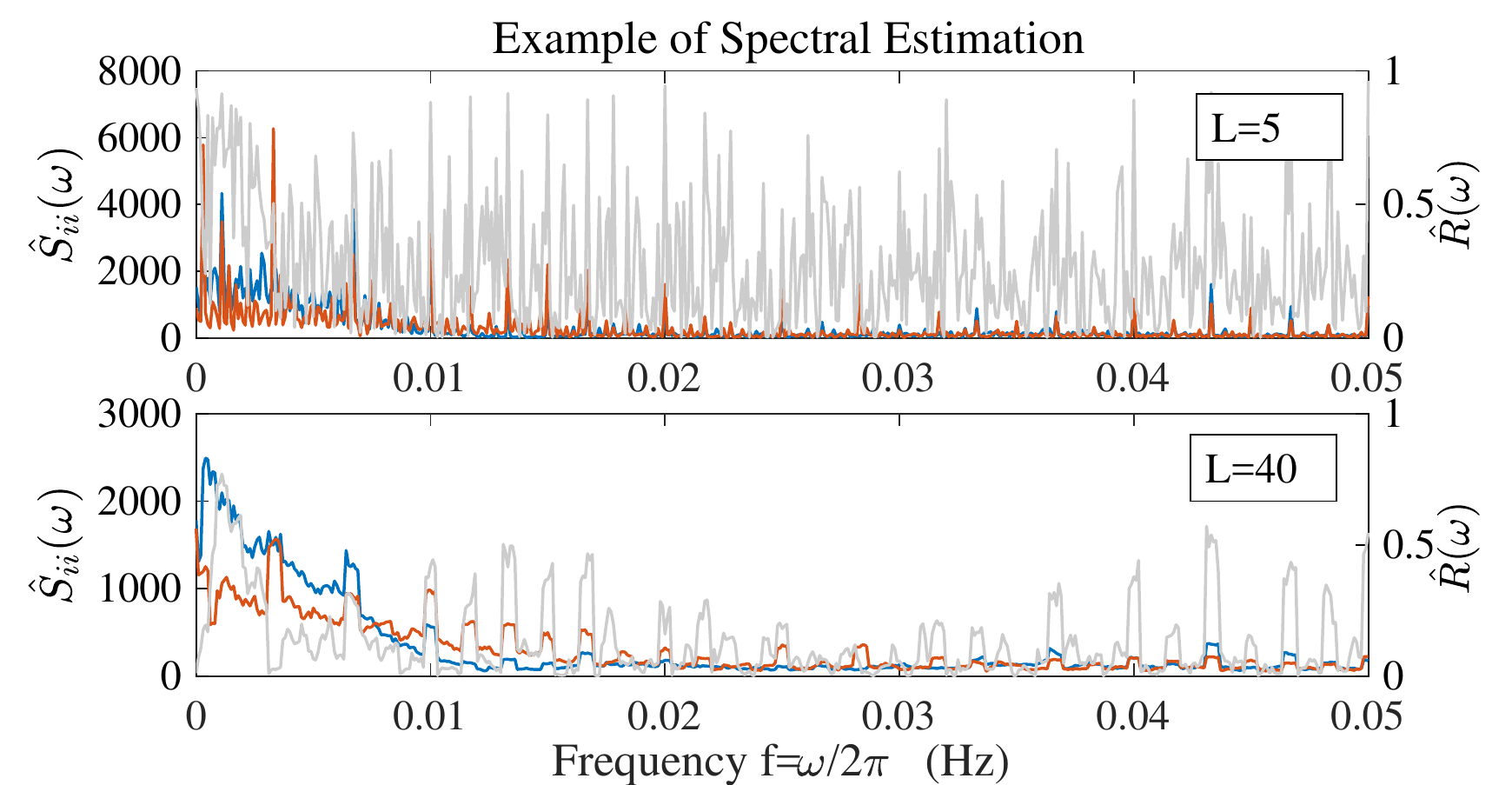}
\par\end{centering}
\caption{Example of estimates for the spectral density. Top: estimation with
$L=5$ tapers. Bottom: estimation with $L=40$ tapers. The red and
blue lines respectively illustrate the on-diagonal spectral density
$\hat{S}_{AB,AB}(\omega)$ and $\hat{S}_{BC,BC}(\omega)$. The grey
line indicates the resultant coherence $\hat{R}(\omega)$.
\label{fig:Example_spectra}}
\end{figure}

\section{Dependency in Network Traffic}

Applying coherence estimation to edge pairs results in a set of estimates
$\{\hat{R}^{(n)}(\omega_{1},\ldots,\omega_{N_{f}})\}$ for $n=1,\ldots,N_{\mathrm{triple}}$.
As may be expected there is significant variation of the spectra across
the set of edge-pairs. In this analysis, we consider fixing the window
of frequencies such that $\omega_{q}=2\pi f_{\max}(q/N_{f})$ for
$q=1,\ldots,N_{f}=500$ and $f_{\max}=0.05Hz$. As the length of the
edge-pair recordings differ, one may desire to increase $L$ as a
function of length $T$. Potentially, this would lead to increased
confidence in our spectral estimate as the Wishart degrees of freedom
are increased. However, such an adaptive tapering scheme where $L$
depends on $T$ creates challenges when comparing across coherence
estimates as it may be hard to disentangle differences due to the
tapering treatment from underlying differences in the process spectra.
As such, in these experiments we decide to fix the number of tapers
at a moderate level $L=40$ for all edge-pairs. The difference between
tapering with $L=5$ and $L=40$ is demonstrated in Figure \ref{fig:Example_spectra}.
Note, that while the cross-spectra for different edge-pairs may be
of a different scale, the coherence (plotted in grey) provides an
intuitive measure on $[0,1]$ allowing comparison across many data-stream
pairs. As an aside, the individual spectra appear non-Poisson, exhibiting shapes that are characteristic of self-exciting behaviour \cite{Hawkes1971}.

\begin{figure}[t]
\includegraphics[width=1\columnwidth]{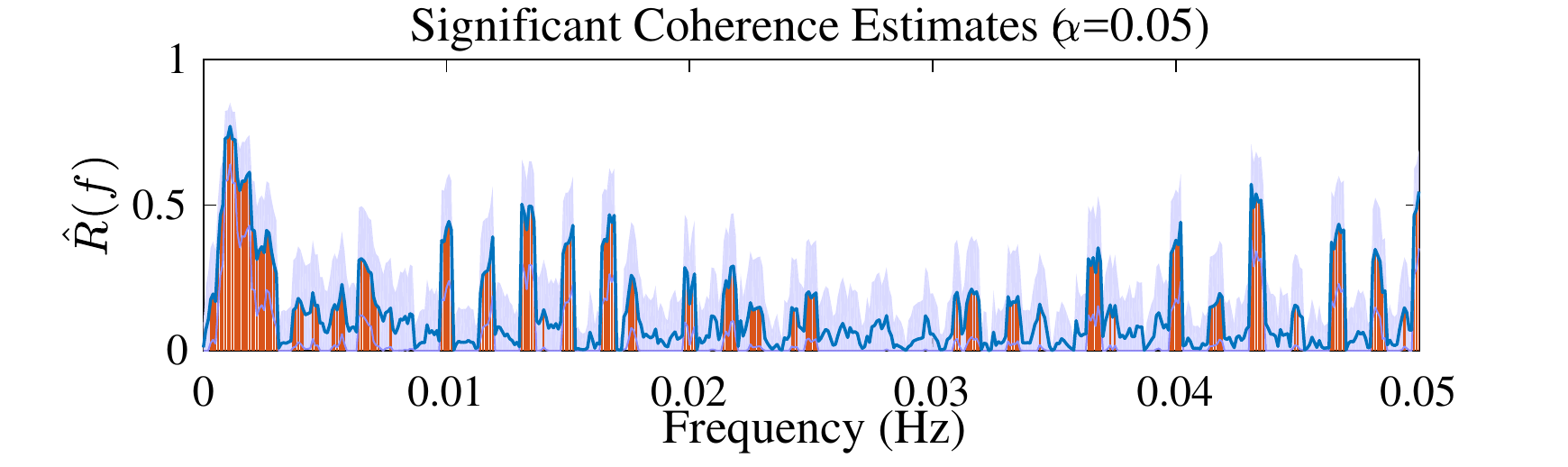}

\caption{The coherence as plotted in the bottom of Fig. \ref{fig:Example_spectra}
with two-sided 95\% confidence intervals. Frequencies where the confidence
interval excludes zero are highlighted in red. \label{fig:significant_coherence}}
\end{figure}

Acknowledging that there will be some error in our spectral estimates,
it is desirable to assign some measure of confidence to estimates.
A useful corollary of the Wishart asymptotic result for multi-taper
estimates is that the coherence is distributed (asymptotically) according
to the Goodman distribution \citep{Carter1987,Goodman1963}. Based
on this distribution, there are a variety of tests that one may perform
to assess the \emph{significance} of a coherence estimate. For example,
one may test the null hypothesis that states $R(\omega_{q})=0$
for each frequency $\omega_{q}$, $q=1,\ldots,N_{f}$. Rather than
test explicitly against a null of zero coherence, in this work we
construct two sided confidence intervals in a similar manner to Wang
et al. \citep{ShouYan2004}. Examples of such intervals for $\alpha=0.05$
are reported in Fig. \ref{fig:significant_coherence}. Alongside these
intervals denoted $[\hat{a}_{\alpha/2},\hat{a}_{1-\alpha/2}]$, we
declare that the coherence at a frequency is significant if the interval
excludes zero. For this particular triple, we note what appears to be significant coherent beaconing across the devices at multiples of approx. 0.0017Hz (a periodicity of 10 minutes). 

To assess variation in coherence estimates across the $N_{\mathrm{triple}}=500$
edge-pairs under study we attempt to cluster the resultant coherence
estimates. Prior to performing clustering, we threshold coherence
estimates according to the confidence intervals such that $\hat{R}^{*}(\omega_{q})=0$
if $0\in[\hat{a}_{\alpha/2},\hat{a}_{1-\alpha/2}]$ and $\hat{R}^{*}(\omega_{q})=\hat{R}(\omega_{q})$
otherwise. The resultant coherence estimates are then modelled as a
\emph{Gaussian mixture model (GMM)}, such that $[\hat{R}^*(\omega_{1}),\ldots,\hat{R}^*(\omega_{N_f})]\sim\mathrm{GMM}(\{\boldsymbol{\mu}_{c},\boldsymbol{\Sigma}_{c}\}_{c=1}^{C})$
where $\boldsymbol{\mu}_{c}\in\mathbb{R}^{N_{f}}$ represent cluster
means and $\boldsymbol{\Sigma}_{c}$ the cluster covariances.

\begin{figure}[t]
\includegraphics[width=1\columnwidth]{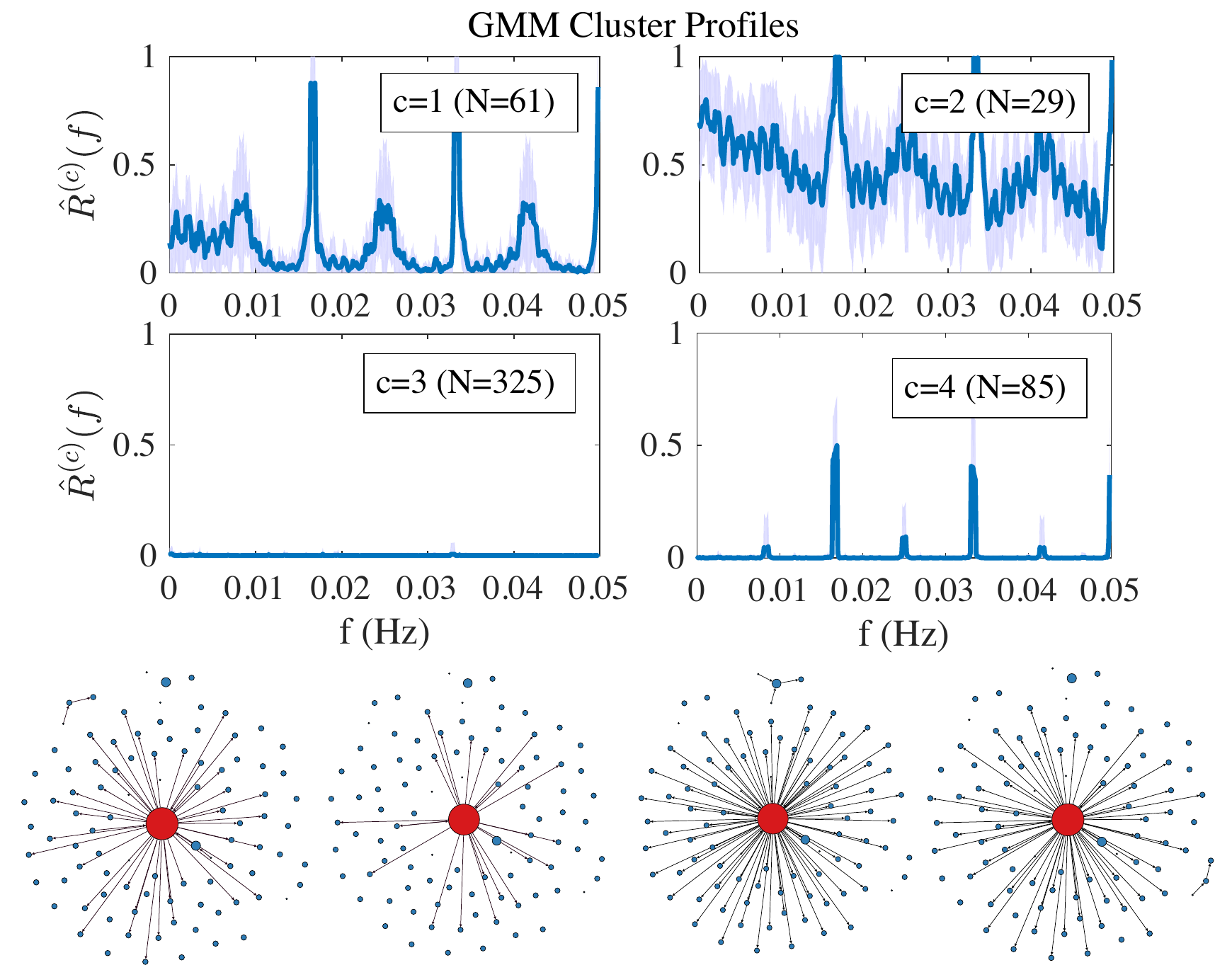}

\caption{GMM clustering of the significance thresholded coherence estimates
with $C=4$. Shaded regions indicate points within one standard-deviation
from the cluster mean. Bottom: graph of edges relating to each cluster
$c=1,2,3,4$ from left to right.\label{fig:K-GMM}}
\end{figure}

While the Gaussian assumption of the above model contrasts with the
Goodman asymptotic distribution for the coherence, the approximation
may still hold relevance. For instance, Enochson and Goodman \citep{Enochson1965}
demonstrated that a Gaussian approximation may be effective when calculating
confidence intervals. In this example we use the MATLAB implementation
of the expectation-maximisation with covariance regularisation set
at $\lambda=0.001$. Due to the many local-minima that may be obtained
when fitting a GMM, we perform one thousand replications with random
initialisation and report the clustering with the largest likelihood.
Figure \ref{fig:K-GMM} presents the resulting mean profiles of
$C=4$ clusters alongside the standard-deviation obtained from the
estimated covariance matrices. 

\section{Discussion}

The clustering results are insightful in that the emergence of clusters
$c=3,4$ partially confirm our initial hypothesis that many device
pairs exhibit little dependency. Out of the initial 500 edge-pairs
90 are placed into clusters with non-negliable coherence, for reference,
the example demonstrated in Fig. \ref{fig:Example_spectra} is placed
into cluster $c=1$. These active coherent clusters exhibit pronounced
structure across multiple frequencies, once again providing evidence
that modelling auto-covariance within network traffic is important.
Of some note, is the clear peak at $f=0.018Hz$ corresponding to periodicity
of around $57$s . Without a more intimate knowledge of device functionality
on the network we can only hypothesise the cause of such a feature,
but it is possible this is due to beaconing activities. Interestingly,
Heard et al. \citep{Heard2014} demonstrate strong periodicity at
this same frequency for devices connecting with dropbox.com. While
most beaconing activities are benign, scanning techniques used by
intruders may also create similar strongly periodic activity and it
is thus of interest to administrators to detect such patterns.

As a direction for future work, it is possible the methods
developed here to detect association between event driven data-streams
in a defensive context, may also be used as a form of correlation
attack, for example to break anonymisation protocols when traffic
is transmitted via mixing devices \citep{Murdoch2007}. It may also
be of interest to relax the stationarity assumptions of this work,
for instance within a wavelet framework. Indeed an algorithm that
could derive the coherence in a streaming manner would be an important
step towards building a practical anomaly detector.

\newpage

\bibliographystyle{ieeetr}
\bibliography{cyber,data}

\end{document}